\documentclass[conference]{IEEEtran}
\IEEEoverridecommandlockouts

\usepackage{listings}
\usepackage{cite}
\usepackage{framed}
\usepackage{lipsum}
\usepackage{multirow}
\usepackage{arydshln}
\usepackage[hidelinks]{hyperref}
\usepackage{amsmath,amssymb,amsfonts}
\usepackage{booktabs}
\usepackage{algorithmic}
\usepackage{graphicx}
\usepackage{textcomp}
\usepackage{xcolor}
\def\BibTeX{{\rm B\kern-.05em{\sc i\kern-.025em b}\kern-.08em
    T\kern-.1667em\lower.7ex\hbox{E}\kern-.125emX}}

\begin{document}

\title{ICST Tool Competition 2025 -- Self-Driving Car Testing Track}


\author{
\IEEEauthorblockN{Christian Birchler\IEEEauthorrefmark{1}\IEEEauthorrefmark{2}, Stefan Klikovits\IEEEauthorrefmark{3}, Mattia Fazzini\IEEEauthorrefmark{4}, Sebastiano Panichella\IEEEauthorrefmark{1}}
\IEEEauthorblockA{\IEEEauthorrefmark{1}\textit{University of Bern, Switzerland}}
\IEEEauthorblockA{\IEEEauthorrefmark{2}\textit{Zurich University of Applied Sciences, Switzerland}}
\IEEEauthorblockA{\IEEEauthorrefmark{3}\textit{Johannes Keppler University Linz, Austria}}
\IEEEauthorblockA{\IEEEauthorrefmark{4}\textit{University of Minnesota, USA}}
}

\maketitle

\begin{abstract}
This is the first edition of the tool competition on testing self-driving cars (SDCs) at the International Conference on Software Testing, Verification and Validation (ICST).
The aim is to provide a platform for software testers to submit their tools addressing the test selection problem for simulation-based testing of SDCs, which is considered an emerging and vital domain.
The competition provides an advanced software platform and representative case studies to ease participants' entry into SDC regression testing, enabling them to develop their initial test generation tools for SDCS.
In this first edition, the competition includes five tools from different authors.
All tools were evaluated using (regression) metrics for test selection as well as compared with a baseline approache.
This paper provides an overview of the competition, detailing its context, framework, participating tools, evaluation methodology, and key findings.
\end{abstract}


\section{Introduction}

Software testing for cyber-physical systems (CPSs) poses different challenges from conventional unit testing.
Many of such challenges lie in the nature of simulation-based testing as simulators are prone to non-determinism and there is a \textit{reality gap}~\cite{DBLP:conf/icst/KhatiriPT23}.
Generally, the greatest limiting factor for using simulators for testing purposes is the cost.
Simulations are costly in terms of time as they require long runtimes and often costly hardware.
Hence, it is vital to use available resources as effectively and efficiently as possible.


In this context, the self-driving car (SDC) Testing Competition offers a platform for software testing researchers to ease their entry into the SDC domain.
The goal of the competition is to motivate them to apply automated testing techniques in a relevant domain: \textit{autonomous vision-based SDC navigation systems}. 
Given the abovementioned context, the competition focuses on determining the tests that are likely to expose a fault in SDCs.
However, the identification of such test cases is challenging and is an open problem in software engineering research.
The identification of test cases is also known as the process of \textit{test selection}.
Test selection is one aspect of the higher level context of \textit{regression testing}, which includes next to test selection also \textit{test minimization} and \textit{test prioritization}~\cite{DBLP:journals/stvr/YooH12}.


Test selection is the process of picking only the relevant test cases from the test suite for a particular change.
In the context of simulation-based testing for SDCs with long-running test cases, we select test cases that reveal as many faults as possible within certain conditions. The competition aims to answer the following research questions (RQs):

\begin{itemize}
    \item \textbf{RQ1:} To what extent do the performances of the tools differ?
    \item \textbf{RQ2:} What are the features leveraged by the top-performing tools?
\end{itemize}

\section{Methodology}
We provided on GitHub an interface specification to the public.
Interested participants in the competition could submit their implementations of the interface as \textit{pull requests}.
We collected the implementations and performed our experiments; we executed them with a benchmark that was not shared with the participants.
To ensure a fair comparison among competing tools and facilitate their development, we have provided participants with an open-source, extensible infrastructure on GitHub:
\begin{center}
    \textit{\small \url{https://github.com/christianbirchler-org/sdc-testing-competition}}
\end{center}

All data are available either on the GitHub repository of the competition or on Zenodo~\cite{zenodo-repo}.

\subsection{Competition Platform}
\begin{figure}[t]
    \centering
    \includegraphics[width=0.6\linewidth]{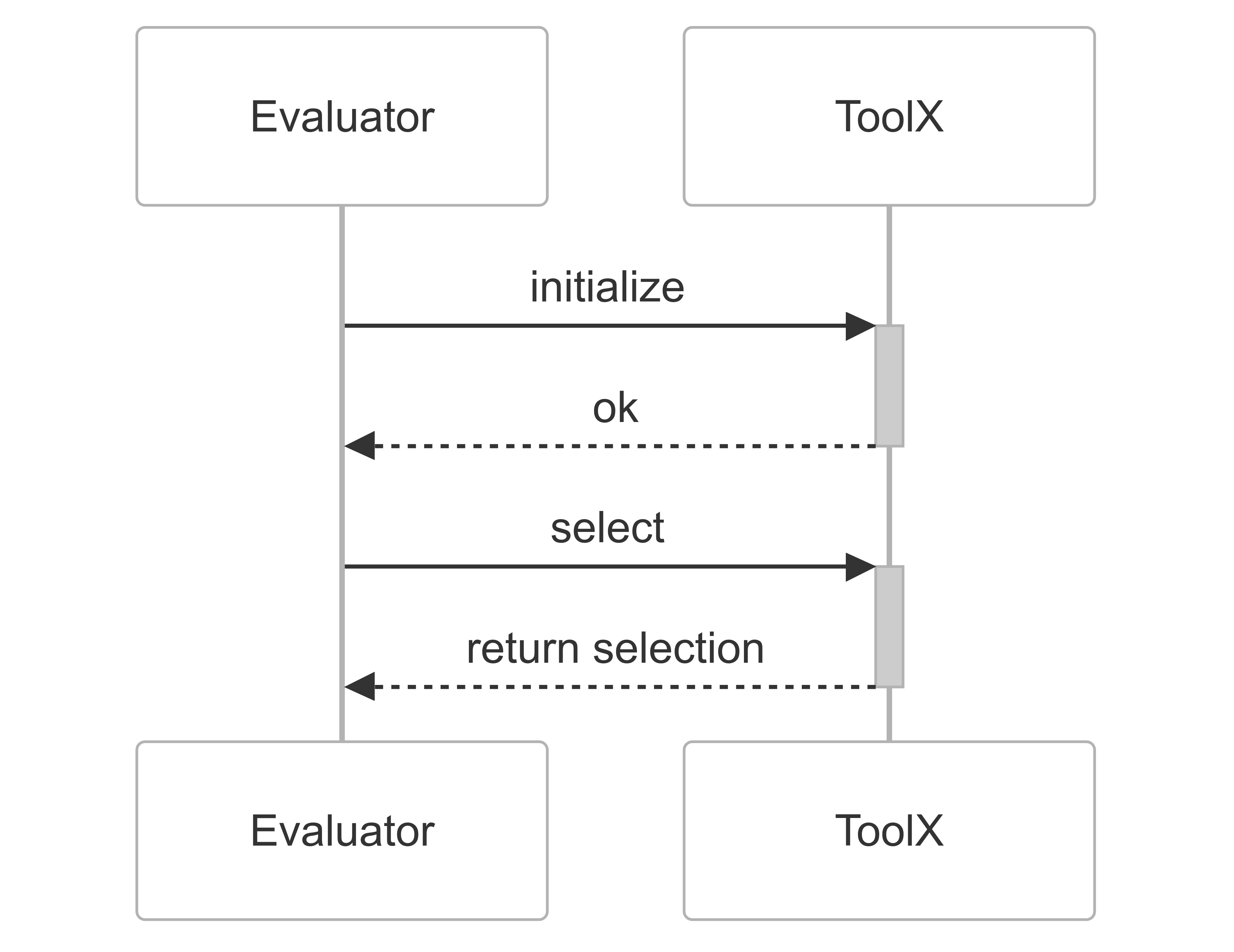}
    \caption{Sequence Diagram of the Tool and Evaluator Intercation}
    \label{fig:sequence-diagram}
\end{figure}

\begin{figure}[t]
    \centering
    \includegraphics[width=0.7\linewidth]{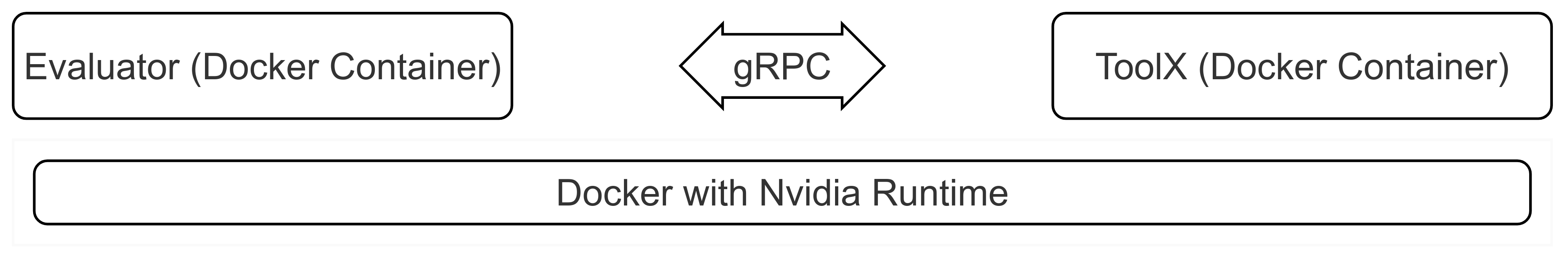}
    \caption{High-level Platform Abstraction}
    \label{fig:platform-overview}
\end{figure}

\subsubsection{Interface}
The participants of the competition had to implement an interface that was provided.
With the interface, interactions such as the tool initialization and selection (see Figure~\ref{fig:sequence-diagram}) are defined.
The interface was defined using the \textit{Protocol Buffers}\footnote{\url{https://protobuf.dev/}} specification\footnote{\url{https://github.com/christianbirchler-org/sdc-testing-competition/blob/main/competition.proto}}.
With \textit{gRPC}\footnote{\url{https://grpc.io/}}, a remote procedure call framework, the user can create implementations of the interface in various programming languages.
For the competition, we aimed to provide the participants as much flexibility as possible.

\subsubsection{Docker}
The flexibility that gRPC provides to the participants implies that the code of the tools is based on different technologies (e.g., Python, Java, etc.).
We mitigate the complexity with the use of Docker.
The participants had to provide, next to their interface implementation also, a docker file, which we used to build images of the tools.

\subsubsection{Sample Tool \& Data}
We also provided an example implementation of the interface on the competition platform as well as an initial set of test cases.
The example implementation is a random test selector, which also acts as a baseline for the evaluation.

\subsection{Competition Tools}
We received five competing tools as overviewed in Table~\ref{tab:tools}.
For each tool, we built a Docker image (x86\_64 architecture) and made them publicly available in the GitHub container registry\footnote{\url{https://github.com/orgs/christianbirchler-org/packages?repo_name=sdc-testing-competition}}.
The details of the tools are explained in the dedicated tool papers of the participants and listed in Table \ref{tab:tools}.

\begin{table}[t]
    \centering
    \scriptsize
    \caption{Overview of Submitted Tools}
    \begin{tabular}{ll}\toprule
         \textbf{Tool}  & \textbf{Docker Image Repository} \\ \midrule
         DRVN Tool      & ghcr.io/christianbirchler-org/drvn\_tool \\ \midrule
         DETOUR         & ghcr.io/christianbirchler-org/detour \\ \midrule
         ITS4SDC        & ghcr.io/christianbirchler-org/its4sdc \\ \midrule
         Graph Selector & ghcr.io/christianbirchler-org/graph\_selector \\ \midrule
         CertiFail      & ghcr.io/christianbirchler-org/certifail \\ \bottomrule
    \end{tabular}
    \label{tab:tools}
\end{table}

\subsection{Benchmark Test Suites}
We use pre-executed test cases based on the SensoDat dataset~\cite{DBLP:conf/msr/BirchlerRKP24}.
SensoDat consists of 36 collections of test cases.
Within a single collection, all test cases are generated and executed under the same conditions, i.e., test generator, driving AI setting, Oracle definition.
Each collection represents a test suite; hence, we have a sample size of 36.
In Table~\ref{tab:benchmarks}, the benchmarks are listed, which are generated by three different test generators.

\subsubsection{Ambiegen}
Humeniuk et al.~\cite{DBLP:conf/sbst/HumeniukAK22} proposed at SBST 2022~\cite{DBLP:conf/sbst/GambiJRZ22} the Ambiegen test generator tool for simulation-based testing of SDCs.
The generator uses a two-objective NSGA-II algorithm to generate the simulation-based test cases.

\subsubsection{Frenetic}
Castellano et al.~\cite{DBLP:conf/sbst/CastellanoCTKZA21} developed for the SBST 2021 tool competition~\cite{DBLP:conf/sbst/PanichellaGZR21} the Frenetic test generator.
Frenetic is also an approach that uses a genetic algorithm with a curvature representation of the road.

\subsubsection{FreneticV}
For the SBST 2022 tool competition, Castellano et al.~\cite{DBLP:conf/sbst/CastellanoKCA22} proposed the FreneticV test generation approach.
It is an extension to the previous Frenetic tool that aims to produce less invalid, i.e., self-intersecting roads.

\begin{table}[t]
    \centering
    \scriptsize
    \caption{Benchmark Overview of SensoDat~\cite{DBLP:conf/msr/BirchlerRKP24}}
    \begin{tabular}{lc} \toprule
        \textbf{Collection (Test Suite)} & \textbf{\# Test Cases} \\ \midrule
        campaign\_2\_ambiegen & 973  \\
        campaign\_2\_frenetic & 928  \\
        campaign\_2\_frenetic\_v & 944  \\
        campaign\_3\_ambiegen & 964  \\
        campaign\_3\_frenetic & 954  \\
        campaign\_4\_ambiegen & 965  \\
        campaign\_4\_frenetic & 964  \\
        campaign\_4\_frenetic\_v & 525  \\
        campaign\_5\_ambiegen & 958  \\
        campaign\_5\_frenetic & 945  \\
        campaign\_5\_frenetic\_v & 940  \\
        campaign\_6\_ambiegen & 959  \\
        campaign\_6\_frenetic & 944  \\
        campaign\_6\_frenetic\_v & 764  \\
        campaign\_7\_ambiegen & 963  \\
        campaign\_7\_frenetic & 967  \\
        campaign\_7\_frenetic\_v & 47  \\
        campaign\_8\_ambiegen & 952  \\
        campaign\_8\_frenetic & 952  \\
        campaign\_9\_ambiegen & 953  \\
        campaign\_9\_frenetic & 964  \\
        campaign\_10\_ambiegen & 971  \\
        campaign\_11\_ambiegen & 973  \\
        campaign\_11\_frenetic & 866  \\
        campaign\_11\_frenetic\_v & 953  \\
        campaign\_12\_frenetic & 956  \\
        campaign\_12\_freneticV & 942  \\
        campaign\_13\_ambiegen & 954  \\
        campaign\_13\_frenetic & 959  \\
        campaign\_13\_frenetic\_v & 951  \\
        campaign\_14\_ambiegen & 959  \\
        campaign\_14\_frenetic & 866  \\
        campaign\_14\_frenetic\_v & 934  \\
        campaign\_15\_ambiegen & 952  \\
        campaign\_15\_frenetic & 870  \\
        campaign\_15\_freneticV & 949 \\ \midrule
        \textbf{Total} & \textbf{32,580}\\ \bottomrule
    \end{tabular}
    \label{tab:benchmarks}
\end{table}

\subsection{Evaluation Metrics}\label{sec:metrics}
We assessed the performance of the tools based on several rather simple metrics.
For the initial edition of this competition, we think those metrics are a reasonable choice.

\subsubsection{Selection Count}
The selection count is the number of test cases selected by a tool.

\subsubsection{Initialization Time}
Every tool gets initial data before the evaluation starts on a separate data set.
The tools can use this initialization phase to train their models (if they use any). 
The initialization time is the time the tool has required for this phase.

\subsubsection{Selection Time}
The selection time is the duration in which a tool performs the selection.
This duration does not include the initialization time.

\subsubsection{Simulation Time to Fault Ratio}
The simulation time-to-fault ratio indicates the cost-effectiveness.
Here, we compute the fraction of the simulation time, i.e., the test execution time, and the number of faults, i.e., the number of failing test cases.
The lower the measure, the better the cost-effectiveness of the tool.

\subsubsection{Fault to Selection Ratio}
The fault-to-selection ratio is another cost-effectiveness metric.
We take the fraction of the number of faults, i.e., the number of failing test cases, and the total actual number of selected test cases by the tool.
This metric can also be seen as the precision of the classification.

\subsubsection{Curvature Diversity}
In testing, especially black-box testing, the test diversity plays an important role~\cite{DBLP:journals/tse/AghababaeyanABS23}.
For the SDC testing context, we define the diversity of the selection based on the roads' curvature profile.
Concretely, for each selected test case, we compute the approximated curvature profile and return its mean value.

\subsection{Experimental Setup}
The experiments are conducted on an Open Stack virtual machine (VM) with 16GB of RAM, eight virtual CPUs, and an Nvidia Tesla T4 GPU.
On the VM is the Docker engine installed on which the tools will run as they are provided as images.
Furthermore, we also used the Nvidia Runtime extension for the docker engine to enable the use of GPU for the tools (see Figure~\ref{fig:platform-overview}).

\subsection{Procedure}
We used each testing campaign from the SensoDat dataset as a sample, and all tools had to select the test cases from each testing campaign.
We consider each testing campaign from the SensoDat dataset as a whole test suite.

All tools treat every test suite independently.
Concretely, for all 36 collections in the SensoDat dataset, we used 80\% for the initialization phase of the tools and the remaining 20\% we used for the evaluation.

\subsection{Data Collection \& Analysis}
The tool evaluator collects the logs and computes the metrics as described in Section~\ref{sec:metrics} for each test suite. Therefore, we obtained for each tool 36 entries for all metrics (as the sample size $N=36$).
For the final evaluation, we consider only the basic statistics of the sample, i.e., the maximum, minimum, mean, and standard deviation.

\section{Experiments and Results}
We ran each tool on all benchmarks, i.e., 36 collections of SensoDat representing a test suite each.
Hence, our sample size is $N=36$.
In Table~\ref{tab:result-metrics}, we computed the statistics of the sample after we ``treated'' them with the participants' tools.
For the sake of comparison, we also provide the statistics of the example tool (provided in the competition platform), which acts as a random baseline since it selects test cases at random ($p=0.5$).

\begin{table*}[t]
    \centering
    \scriptsize
    \caption{Performance Metrics Statistics ($N=36$)}
    \begin{tabular}{cccccccc} \toprule
         \multirow{2}{*}{\textbf{Tool}}     & \multirow{2}{*}{\textbf{Statistic}}  & \multicolumn{6}{c}{\textbf{Metrics}}  \\
                                            &   & selection\_cnt & time\_to\_initialize & time\_to\_select\_tests & time\_to\_fault\_ration & fault\_to\_selection\_ration & diversity \\ \midrule
         \multirow{4}{*}{Random}            & max & 107.000000 & 0.344903 & 0.097950 & 297.940525 & 0.561224 & 0.068561 \\
                                            & mean      & \textbf{88.166667}     & 0.269580 & 0.076621 & \textbf{156.231409} & \textbf{0.379028} & 0.039256 \\ 
                                            & std  & 18.673893 & 0.050076 & 0.014376 & 92.451823 & 0.081674 & 0.011048 \\ 
                                            & min  & 4.000000 & 0.026098 & 0.008159 & 18.243495 & 0.221053 & 0.026567 \\ \hdashline
         \multirow{4}{*}{DRVN Tool}         & max  & 35.000000 & 1.948507 & 19.937558 & 138.899388 & 1.000000 & 0.057320 \\ 
                                            & mean      & \textbf{14.000000}     & 0.559066 & 10.485846 & 72.957767 & 0.742443 & 0.039737 \\
                                            & std  & 9.276886 & 0.252115 & 3.434871 & 40.181381 & 0.137023 & 0.012785 \\
                                            & min & 1.000000 & 0.462480 & 7.394846 & 8.482704 & 0.357143 & 0.000000 \\ \hdashline
         \multirow{4}{*}{DETOUR}            & max       & 78.000000     & 66.941565 & 17.080501 & 243.457197 & 0.750000 & 0.068686 \\
                                            & mean  & 68.361111 & \textbf{61.773342} & \textbf{15.668664} & 124.372055 & 0.465910 & 0.038079 \\
                                            & std  & 17.284432 & 11.653761 & 2.928543 & 71.568387 & 0.107662 & 0.012462 \\
                                            & min  & 4.000000 & 3.195855 & 0.861387 & 16.525530 & 0.283784 & 0.024868 \\ \hdashline
         \multirow{4}{*}{ITS4SDC}           & max & 99.000000 & 0.386055 & 1.136053 & 109.066821 & 1.000000 & 0.065439 \\
                                            & mean & 67.805556 & 0.297533 & 0.966264 & \textbf{65.390933} & \textbf{0.804806} & 0.036470 \\
                                            & std & 16.763599 & 0.056466 & 0.209256 & 33.464099 & 0.060595 & 0.010112 \\
                                            & min & 4.000000 & 0.026640 & 0.075265 & 11.913698 & 0.691358 & 0.022947 \\ \hdashline
         \multirow{4}{*}{CertiFail}         & max & 78.000000 & 0.373093 & 3.067581 & 366.407264 & 0.744681 & 0.052254 \\
                                            & mean & 47.722222 & \textbf{0.273155} & 2.788512 & 129.452883 & 0.515544 & 0.030239 \\
                                            & std & 14.328017 & 0.055594 & 0.515386 & 79.790441 & 0.094049 & 0.008873 \\
                                            & min & 4.000000 & 0.024092 & 0.169545 & 18.284210 & 0.250000 & 0.009570 \\ \hdashline
         \multirow{4}{*}{Graph Selector}    & max & 78.000000 & 1.040956 & 0.243248 & 215.126609 & 0.684211 & 0.055859 \\
                                            & mean & 72.305556 & 0.447819 & \textbf{0.211349} & 115.918430 & 0.535351 & 0.034583 \\
                                            & std & 12.448338 & 0.121785 & 0.038682 & 65.053108 & 0.079143 & 0.008917 \\
                                            & min & 10.000000 & 0.140035 & 0.015087 & 17.290175 & 0.391892 & 0.023591 \\ \bottomrule
    \end{tabular}
    \label{tab:result-metrics}
\end{table*}

\begin{framed}
\small 
\textbf{Finding 1:}
On average, the random test selector selects the most test cases ($\mu = 88.17$) compared to all tools.
\end{framed}

\begin{framed}
\small 
\textbf{Finding 2:}
DETOUR required on average ($\mu = 61.77 sec$) to initialize the tool with the provided initialization data.
All other tools did not require more than a second.
\end{framed}

\begin{framed}
\small 
\textbf{Finding 3:}
For the selection phase, DETOUR requires, on average, most of the time ($\mu = 15.67 sec$) followed by DRVN Tool ($\mu = 10.49 sec$).
The remaining tools did not require more than three seconds on average.
\end{framed}

\begin{framed}
\small 
\textbf{Finding 4:}
The random selector has, on average, the worst \textit{Simulation Time to Fault Ratio} with $\mu = 156.23$.
With an average of $\mu=65.39$, the ITS4SDC tool has the best ratio.
\end{framed}

\begin{framed}
\small 
\textbf{Finding 5:}
The ITS4SDC tool has, on average, the best \textit{Fault to Selection Ratio} ($\mu = 0.8$) and the random baseline the worst with $\mu = 0.38$.
\end{framed}

\begin{framed}
\small 
\textbf{Finding 6:}
There is no clear indication that the diversity metric differs among all tools.
\end{framed}

\section{Discussion}
The random test selector selects more test cases than all other tools.
With this observation, we have an indication that the competing tools leverage different strategies to select fewer test cases but with the aim of having only failing ones.
Hence, no random approach is involved but most likely models, which predict the test outcomes.

DETOUR needs among the tools the most time to initialize.
The tool has a model implementation that requires a costly training phase.
However, depending on the time budget available for a testing campaign, the required initialization time might be negligible from a practical point of view.

DETOUR also needs the most time among the tools for the selection followed by the DRVN Tool.
As mentioned above, DETOUR needs time for the initialization, but the DRVN Tool does not even need a second.
This can be explained because the DRVN Tool comes already with a pre-trained model, hence not much initialization for training is required.
If this is said, the high selection time for both tools can be an indication that their models are rather complex, inducing a slower selection performance.

ITS4SDC identifies the most faults related to the simulation time of the selected test cases.
Regarding cost-effectiveness, where cost relates to the testing time and effectiveness in the identification of fault-revealing test cases, the ITS4SDC tool has the best performance.
Furthermore, when cost-effectiveness relates to the number of test cases in the selection and the number of fault-revealing test cases, ITS4SDC also shows the best performance.


\section{Conclusion}
Test selection for simulation-based test suites remains an open problem.
This year marks the first edition of the SDC Testing Competition.
We evaluated and compared five tools with a random approach as a baseline.
According to our results, the best-performing tool is ITS4SDC.
In this competition edition, we introduced evaluation metrics aimed at assessing how effectively the tools submitted select only the failing test cases. 
For the upcoming years, we envisage introducing other kinds of metrics \cite{BirchlerM0NKP24}, obstacles in the scene (e.g., trees), and other environmental factors, including weather conditions.

\section*{Acknowledgments}
We thank the participants
of the competition for their invaluable
contribution. 
We thank 
the Horizon 2020 (EU Commission) support for the \href{https://www.innoguard.eu/index.html}{project InnoGuard}, 
Marie Sktodowska-Curie Actions Doctoral Networks (HORIZON-MSCA-2023-DN),  and the SNSF project entitled "SwarmOps: Human-sensing based MLOps for Collaborative Cyber-physical systems" (Project  No. 200021\_219732).
Furthermore, we also thank CHOOSE, the Swiss Group for Original and Outside-the-box Software Engineering, for their financial support.

\bibliographystyle{IEEEtran}
\bibliography{main}

\end{document}